\documentclass[a4paper]{article}

\usepackage{spconf, graphicx, booktabs}
\usepackage{tabularx} % in the preamble
\usepackage{array}
\usepackage{multirow}
\usepackage{hyperref}
\usepackage{cite}
\usepackage[shortlabels]{enumitem}
\setlist[enumerate]{nosep}

\hypersetup{
    colorlinks=true,
    linkcolor=black,
    allcolors=black,
    filecolor=black,      
    urlcolor=blue,
}
\usepackage{enumitem}

\title{Dynamic Layer Customization for Noise Robust Speech Emotion Recognition in Heterogeneous Condition Training}
\name {Alex Wilf, Emily Mower Provost}
\address{Computer Science and Engineering\\
University of Michigan, Ann Arbor, Michigan, USA}

\begin{document}
\maketitle

\begin{abstract}
Robustness to environmental noise is important to creating automatic speech emotion recognition systems that are deployable in the real world. Prior work on noise robustness has assumed that systems would not make use of sample-by-sample training noise conditions, or that they would have access to unlabelled testing data to generalize across noise conditions. We avoid these assumptions and introduce the resulting task as heterogeneous condition training. We show that with full knowledge of the test noise conditions, we can improve performance by dynamically routing samples to specialized feature encoders for each noise condition, and with partial knowledge, we can use known noise conditions and domain adaptation algorithms to train systems that generalize well to unseen noise conditions.  We then extend these improvements to the multimodal setting by dynamically routing samples to maintain temporal ordering, resulting in significant improvements over approaches that do not specialize or generalize based on noise type.
\end{abstract}


\section{Introduction}
\label{sec: intro}
Automatic emotion recognition provides an opportunity to understand how emotion patterns in daily life are associated with health, both mental and physical \cite{law2020automatic, beale2008role}.  The inexpensive production of audio recording-capable devices has made speech emotion recognition (SER) an attractive avenue for the deployment of emotion recognition systems and recent advances in machine learning have led to improved accuracy in state-of-the-art SER systems, but robustness to additive noise in SER is still an open problem.  In this work, we introduce a novel task for training and testing SER systems for noise robustness that simulates real-world use; demonstrate that we can successfully customize feature encoders to noise conditions known at training time; apply domain adaptation methods commonly used to generalize performance across datasets to the task of generalizing across noise conditions; and extend these improvements to the multimodal setting using a process we describe as Dynamic Layer Customization (DLC).

In considering how a SER system bound for real-world deployment would be developed, it is reasonable to assume that the system's designers may have some knowledge of the noise conditions that will appear at deployment, either through empirical studies or expert knowledge.  Prior works on noise robust speech tasks have either assumed that when a system trains with all test noise conditions known during training, the system makes no use of the noise conditions on a sample-by-sample basis, or when a system trains on a subset of the test noise conditions, it is able to use the unlabelled samples noised with the ``unseen" noise conditions in training \cite{eyben2012improving, liao2018noise, meng2017unsupervised}.  In this paper, we introduce a novel noise robustness task, \emph{heterogeneous condition training}, where systems either have access to all or some of the test noise conditions on a sample-by-sample basis during training, and ``unseen" noise conditions remain unseen during training.  We show, for the first time, that networks which dynamically route samples based on noise condition can enhance noise robust SER performance when training noise conditions match testing noise conditions, and that domain adaptation can be used without unlabelled target data to improve performance when some noise conditions are unseen in training.

Our first experiment examines the case where a network has access to all test noise conditions at training, and shows that it is possible to specialize ``expert" feature encoders for each noise condition -- which we call \emph{full customization} -- and improve performance over the same system implemented with a single feature encoder. The situation where a system designer has access to all test noise conditions at train time is justified in a deployment scenario in which a small number noise conditions make up such a majority of cases or a certain noise condition is so frequent as to justify a system designed with a noise condition predictor and a specialized feature encoder to handle it.  This is a common paradigm for technology deployed in a static environment, for example, a single clinic.

A simple implementation of full customization would be to split the dataset into subdatasets by noise type and train and test networks with separate feature encoders on each subdataset.  This approach, which we will refer to as \emph{statically} customizing layers, introduces a subtle yet insidious problem: splitting the dataset corrupts the temporal order of the samples when neighboring samples contain varying types of noise -- a common phenomenon in natural collections of emotional data \cite{zhang2016facing}.  While unimodal acoustic SER models usually consider each utterance independently, multimodal models frequently require temporal consistency, as they use context from neighboring samples to predict the emotional content of each sample \cite{poria2017context, mai2019divide}. To solve this problem, we introduce a novel paradigm in noise robust SER: dynamically routing samples to different feature encoders based on noise condition and recombining the outputs in the original order through DLC. Full customization can be thought of as a special case of a mixture-of-experts (MoE) model \cite{shazeer2017outrageously}, in which the number of experts in each ensemble is one. 
% Dynamic adjustment to noise is not entirely new to speech tasks; Kim et al. \cite{kim2016environmental} dynamically adjusted to noise by concatenating intermediate features related to the noise present in the sample to the intermediate feature representations used for classification on the task of speech recognition.  Our approach differs from this in that we dynamically adjust the path a sample takes through the network as opposed to adding to the representation of the partially processed sample.

Our second experiment examines the case where a network has access to some, but not all, test noise conditions at training. We support the finding, from related speech tasks, that domain adaptation can be used to generalize across noise conditions \cite{liao2018noise, meng2017unsupervised}, and extend these findings to the case where unlabelled target information is not provided to the network during training.  We show that domain adaptation methods can lead to significant performance improvements over single and fully customized networks.  We choose the highest performing domain adaptation method from our unimodal tests, Domain Separation Network (DSN), to extend to the multimodal setting.  As DSN uses separate feature encoders for each domain, we use DLC to extend DSN to the multimodal setting so as not to disrupt the temporal ordering of samples.

In rest of this paper, we detail the methods, data, and experiments used and show that specialized feature encoders can improve emotion recognition in the presence of known noise conditions; domain adaptation can be used without unlabelled target data to generalize to unseen noise conditions; and dynamic routing through DLC can extend these performance improvements to the multimodal setting.

\section{Methods}
\label{sec: methods}
Our baseline network consists of a feature encoder layer linked to both an emotion classifier (trained with cross entropy loss) and a decoder (trained with mean squared error loss, comparing the output with the clean input to encourage denoising).  The feature encoder layer either contains a single feature or encoder or multiple (in the cases of full customization and DSN). Our architectures for each component (feature encoder, classifier, decoder, and adversary) are constant across all networks, and are based on Khorram et al.'s approach to unimodal acoustic emotion recognition using dilated convolutions \cite{khorram2017capturing}.

The \textbf{feature encoder} is implemented with three 1-D convolution layers with kernel size 16, 128 feature maps, and dilation rates increasing by powers of two with each successive layer as in \cite{khorram2017capturing, gideon2019improving}, followed by a 1-D MaxPool with pool size 4 and 4 strides.  The \textbf{decoder layer} for feature reconstruction consists of two 1-D convolution layers with 128 and 40 feature maps, kernel size 3, and 2 strides, followed by a single 1-D convolution layer with 40 feature maps, kernel size 3, and 1 stride.  The \textbf{adversary} and \textbf{classifier} layers each consist of three dense layers with two 128 unit layers followed by a layer where the number of units is either the number of noise conditions (adversary) or the number of emotion bins (classifier).  For each method, we use the Adadelta optimizer with learning rate $1\mathrm{e}{-3}$.

We test unimodal and multimodal variations of this architecture that leverage customization and domain adaptation for heterogeneous condition training.  In the subsections that follow, we describe the details of the domain adaptation methods we test, DLC, and the multimodal setting of the task.

\subsection{DANN}
\label{subsec: dann}
The Domain Adversarial Neural Network \cite{ganin2015unsupervised} is an approach to domain adaptation in which features are passed through a feature encoder, then the encoded features are passed through a task classifier and an adversarial domain classifier (the latter is preceded by a gradient reversal layer).  In this way, the feature encoder is encouraged to output encodings such that using those encodings, the task classifier is able to predict the task, but the domain classifier is unable to predict the domain. DANN has shown promising results on cross-corpus vision tasks, and has been successfully applied to SER, though the authors noted that it had difficulty converging \cite{abdelwahab2018domain}.

\subsection{MADDoG}
\label{subsec: maddog}
The Multiclass Adversarial Discriminative Domain Generalization network \cite{gideon2019improving} is a variation on DANN where the adversary (called a ``critic") uses a linear activation with loss based on WGAN-style ``earth mover's distance" \cite{arjovsky2017wasserstein} instead of cross-entropy loss with a softmax activation.  The critic is trained separately at the beginning of each epoch and then is frozen.  MADDoG has shown promising results on the task of SER in domain generalization -- a domain adaptation variation where some labelled test samples are available at training.

\subsection{DSN}
\label{subsec: DSN}
The purpose of the Domain Separation Network (DSN) \cite{bousmalis2016domain} is to learn a ``shared" encoder that extracts features that are \emph{generalizable} across domains.  DSN achieves this by learning ``private" feature encoders that encode the parts of a sample \emph{unique} to each domain, trained with losses to encourage that, for each sample: the shared and private encoders yield different representations; all information relevant to reconstruction is captured by the normalized sum of the outputs of the shared and private encoders; the shared encoding is sufficient to classify the task; and the outputs of the shared encoder for samples from different domains are so similar as to be indistinguishable by a DANN-style adversary.  In the original paper, DSN created batches by randomly sampling from each domain, but as doing so disrupts the original ordering we require to extend to the multimodal setting (described in Section~\ref{subsec: multimodal}), we use DLC to dynamically route samples to different private encoders and back for reconstruction and classification.  

\subsection{Multimodal Setting}
\label{subsec: multimodal}
We extract transcripts over the noise enhanced audio files using Deepspeech \cite{hannun2014deep} and pre-process the transcripts using BERT embeddings \cite{devlin2018bert}.  We use a state-of-the-art multimodal fusion network: Hierarchical Feature Fusion Network (HFFN) \cite{mai2019divide}.  HFFN extracts independent features for each modality (i.e., lexical, acoustic) before ``fusing" them and learning to recognize emotion using cross-utterance context. We refer the reader to Mai et al. \cite{mai2019divide} for additional details.  

We use DLC to test different methods as unimodal acoustic feature encoders for HFFN, maintaining the original ordering HFFN requires for cross-utterance learning.  To do this, we split each batch of input samples into sub-batches based on noise condition. We process each sub-batch through its respective feature encoder, then recombine samples using their original indices in the batch before passing the encodings on to the rest of the network.  Doing so enables a unimodal feature encoder layer to use multiple specialized experts, either for full customization or for domain adaptation (in the case of DSN), as part of an end-to-end multimodal network.
%For example, a batch with three samples (s) and 2 noise conditions (nc) known at train time could be of the form [(s1,nc1), (s2,nc2), (s3,nc1), (s4,nc2)].  DLC would first split this batch into sub-batches [(s1,nc1), (s3,nc1)] and [(s2,nc2), (s4,nc2)], remembering that s1 came from index 1 in the original batch, s2 came from index 2...etc.  We would pass each sub-batch through its respective feature encoder, then reform the outputs into the original batch form, this time with outputted encodings (e) from the individual feature encoders: [(e1, nc1), (e2, nc2), (e3, nc1), (e4,nc2)].

\begin{table*}[ht]
\caption{UAR results for all methods on both experiments across datasets. Italics indicate a multimodal setting.}
\label{tab: results}
\footnotesize
\centering
\begin{tabular}{ll|lll|lll|lll}
                %   \cline{3-11}
 \toprule
                       & \textbf{Dataset}       & \multicolumn{3}{c|}{\textbf{MSP}} & \multicolumn{3}{c|}{\textbf{IEMOCAP}} & \multicolumn{3}{c}{\textbf{MOSEI}}                                                  \\
                  
                       & \textbf{Noise Profile} & h1                               & h2                              & h3                                 & h1           & h2    & h3    & h1    & h2    & h3    \\
                            \hline
 \textbf{Experiment 1} & none                   & 48.95                            & 51.17                           & 51.59                              & 55.14        & 56.49 & 57.29 & 39.26 & 39.42 & 40.13 \\
                       & single                 & 51.39                            & 53.51                           & 54.99                              & 58.29        & 59.01 & 60.55 & 42.25 & 42.98 & 43.40  \\
                       & multi                  & 53.05                            & 55.10                            & 56.95                              & 60.21        & 62.41 & 62.45 & 44.50  & 44.75 & 45.61 \\
                       & multi-DLC              & 53.12                            & 54.97                           & 56.89                              & 60.36        & 62.23 & 62.70  & 44.74 & 44.86 & 45.72 \\
                       & \emph{HFFN-single}            & 57.47                            & 59.75                           & 60.68                              & 63.98        & 66.17 & 66.88 & 49.41 & 48.97 & 50.22 \\
                       & \emph{HFFN-multi-DLC}         & \textbf{58.98}                            & \textbf{61.05}                           & \textbf{62.79}                              & \textbf{65.37}        & \textbf{67.96} & \textbf{68.48} & \textbf{51.44} & \textbf{50.19} & \textbf{51.97} \\
                            \hline
 \textbf{Experiment 2} & none                   & 48.45                            & 49.58                           & 50.18                              & 53.25        & 53.99 & 55.67 & 37.17 & 37.82 & 37.63 \\
                       & single                 & 49.58                            & 51.28                           & 51.70                               & 55.81        & 56.94 & 57.92 & 39.38 & 40.53 & 40.91 \\
                       & multi                  & 48.21                            & 49.78                           & 49.16                              & 52.24        & 51.43 & 54.25 & 38.21 & 39.08 & 39.44 \\
                       & multi-DLC              & 48.16                            & 49.96                           & 49.21                              & 52.18        & 51.43 & 54.29 & 38.19 & 39.12 & 39.41 \\
                       & DANN                   & 51.50                             & 52.10                            & 52.29                              & 57.40         & 58.26 & 59.80  & 40.15 & 41.23 & 42.18 \\
                       & MADDoG                 & 51.88                            & 52.88                           & 52.37                              & 58.16        & 58.28 & 60.48 & 40.19 & 41.75 & 42.74 \\
                       & DSN-DLC                & 52.22                            & 53.13                           & 54.44                              & 57.92        & 60.12 & 61.61 & 41.21 & 42.50  & 43.32 \\
                       & \emph{HFFN-single}            & 55.83                            & 57.29                           & 58.53                              & 62.70         & 64.83 & 66.02 & 45.62 & 46.66 & 47.23 \\
                       & \emph{HFFN-dsn-DLC}           & \textbf{58.28}                            & \textbf{59.31}                           & \textbf{60.66}                              & \textbf{66.08}        & \textbf{66.05} & \textbf{67.90}  & \textbf{46.66} & \textbf{48.48} & \textbf{49.83} \\

\bottomrule
\end{tabular}
\end{table*}
                    
\section{Data}
\label{sec: data}

\subsection{Datasets}
We consider three datasets in our experiments. \textbf{MSP-Improv} is an acted, audiovisual emotional database which aims to approach the naturalness of unsolicited human interactions by asking the actors to embed a ``target sentence" into an improvised interaction \cite{busso2016msp}. MSP-Improv was collected over six sessions with twelve actors and contains 8,438 utterances, each labelled for valence and activation on a scale of 1-5. We convert the $n$ valence ratings into a three bin vector describing the distribution of the sample over ``low", ``medium", and ``high" valences by binning ratings below, equal to, and greater than the midpoint (3) and dividing by $n$ as in \cite{gideon2019improving}.  \textbf{MOSEI} contains 23,500 utterances extracted from ``in the wild" videos on Youtube, labelled for sentiment in the range -3 to 3 \cite{zadeh2018multimodal}. We also consider negative, neutral, and positive bins for MOSEI, this time by partitioning ratings with 0 as the midpoint. \textbf{IEMOCAP} was collected over five sessions from ten actors (five male, five female) \cite{busso2008iemocap}. Each of the 10,039 utterances is labelled with emotional categories (e.g., anger, happiness, sadness, neutrality) and dimensional labels (i.e., valence, activation, dominance).  Though we use dimensional labels to evaluate results on MSP-Improv and MOSEI, we evaluate performance on IEMOCAP using categorical labels to be consistent with prior work \cite{mai2019divide, poria2017context}.

\subsection{Feature Extraction}
We use the Librosa Python library \cite{mcfee2015librosa} to extract 40 dimensional log Mel Filterbanks (MFB), which have shown effectiveness in SER \cite{khorram2017capturing, gideon2019improving}.  We also use Librosa, along with the ESC-50 environmental noise dataset \cite{piczak2015esc}, to overlay additive noise with different signal to noise ratios (SNR), detailed below. Our experiments were run across a single machine using 3 GPUs: 1x GTX 1080, 2x Titan X.  The code to reproduce our results will be posted by the authors.  

\section{Experiments}
\label{sec: experiments}
In our experiments, we add noise to each dataset in different profiles, selecting randomly from among three noise conditions comprising both a noise type -- either ``natural", ``human", or ``interior" -- and a signal to noise ratio (SNR).  After selecting a noise condition, we overlay the original sample with a randomly chosen audio file from that category of the ESC-50 dataset \cite{piczak2015esc} at the given SNR. Our noise profiles are inspired by real life situations.  h1 is inspired by a grocery store environment, with (natural, interior, human) SNR values of (-5, -20, -20).  h2 is inspired by a sidewalk environment, with values of (-20, -1, -5).  h3 is inspired by an interior environment, with values of (-5, -30, -10).  Each result reported in Table~\ref{tab: results} is the average of five trials.

We need access to the noise condition of samples at test time for our full customization network to be able to route samples to different feature encoders.  However, this is not realistic for a deployed algorithm, so we train a noise predictor to generate noise condition labels for test samples. It is a feature encoder followed by a classifier layer, and averages 87\% accuracy at predicting noise condition labels on the test set (across noise profiles).

\subsection{Experiment 1}
We first consider the case where all test noise conditions are known at train time, and test whether a fully customized network with multiple feature encoders -- one for each noise condition -- will outperform the baseline network (labelled ``single" in Table~\ref{tab: results}).  We examine splitting both statically (labelled ``multi") and dynamically (labelled ``multi-DLC"), and extend our baseline and multi-DLC models to the multimodal setting with HFFN (as extending multi without DLC would corrupt the order of the samples).

\subsection{Experiment 2}
Next, we investigate if performance can be improved in the case where we leave one noise condition out during training (a common testing approach in domain adaptation problems \cite{gideon2019improving, bousmalis2016domain}). In this experiment, we examine whether the domain adaptation methods described in Section~\ref{sec: methods} can encourage the feature encoders to generalize across noise conditions.  We choose the most promising method from our unimodal tests (DSN-DLC) to extend to the multimodal setting, using DLC to maintain the order of samples. The reported performance is the average performance of the model, leaving each noise condition out once as the unseen noise condition.

\section{Results}
\label{sec: results}
Results for both experiments are listed in Table~\ref{tab: results}. In Experiment 1, our results support our hypothesis that a fully customized network with multiple feature encoders specialized for particular noise conditions (multi) outperforms our baseline network with a single feature encoder (single).  In the unimodal setting, the fully customized network implemented dynamically using DLC performs similarly to the same network implemented statically, and shows an average improvement of 2.13\% UAR (unweighted average recall) across noise profiles and datasets over the baseline network.  In the multimodal setting, this difference is less pronounced: 1.63\%.  We hypothesize that this is due to the fact that lexical embeddings play a prominent role in SER, so differences only applied to acoustic features will result in a smaller improvement.  Future work may find more improvements in the multimodal setting by using denoised audio for transcription.

In Experiment 2, we found that the fully customized network with multiple feature encoders did not outperform the baseline with a single feature encoder.  We believe that this is because there is no generalization within the specialized feature encoders, so each encoder is poorly equipped to handle samples noised with the test noise condition.  Our results showed that the domain adaptation methods significantly improved upon the baseline (single) and fully customized (multi) approaches. DSN performed the best, improving on a single, ungeneralized feature encoder by an average of 2.49\%.  We implement DSN (the highest performing method in the unimodal setting) with DLC in the multimodal setting, and find an average improvement of 2.06\% over our multimodal network with the single ungeneralized feature encoder.
% As described in Section~\ref{sec: methods}, this feature encoder architecture is consistent across all methods.

\section{Conclusion}
In this paper, we present heterogeneous condition training as a novel training and evaluation task for noise robustness in SER that permits full or partial knowledge of test noise conditions at training time.  We improve performance on this task over our baseline network by training specialized subnetworks and applying domain adaptation methods in the absence of unlabelled target data, treating noise conditions as domains. We extend these findings to the multimodal setting by dynamically routing samples to and from specialized feature encoders, maintaining the temporal order of the samples.  We believe that heterogeneous condition training provides a useful task for future work in noise robustness to test against, and that our findings -- that individual feature encoders can effectively specialize to specific noise conditions, and that domain adaptation methods can be used to generalize to unseen noise conditions -- will help shape the deployment of SER systems in the real world.

\section{Acknowledgments}
This work was supported by the National Science Foundation (NSF CAREER 1651740) and  Precision Health at the University of Michigan.  Special thanks go to Amir Zadeh, Zakaria Aldeneh, and Katie Matton.

\clearpage
\newcommand{\BIBdecl}{\setlength{\itemsep}{0.25 em}}
\bibliographystyle{IEEEtran}

\bibliography{ms}

\end{document}